\documentclass[twocolumn,10pt,a4paper,superscriptaddress,showpacs,preprintnumbers,amsmath,amssymb,notitlepage,nofootinbib]{revtex4-1}
\date{March 2010}
\usepackage{epsfig}
\usepackage{latexsym}
\usepackage{amssymb}
\usepackage{booktabs}
\usepackage{graphicx}% Include figure files
\newcommand{\be}{\begin{equation}}
\newcommand{\ee}{\end{equation}}
\newcommand{\ba}{\begin{eqnarray}}
\newcommand{\ea}{\end{eqnarray}}
\newcommand{\bi}{\begin{itemize}}
\newcommand{\ei}{\end{itemize}}

\newcommand{\<}{\langle}
\renewcommand{\>}{\rangle}

\newcommand{\la}{\label}

\newcommand{\Deff}{D_{\mathrm{eff}}}

\begin{document}
\preprint{CERN-TH-2020-004, MITP/20-001}
\title{The rate of photon production in the quark-gluon plasma from lattice QCD}

\author{Marco C\`e} 
\affiliation{Helmholtz-Institut Mainz, Johannes Gutenberg-Universit\"at Mainz,
D-55099 Mainz, Germany}
\affiliation{Theoretical Physics Department, CERN, CH-1211 Geneva 23, Switzerland}

\author{Tim Harris}
\affiliation{Dipartimento di Fisica, Universit\`a
 di Milano–Bicocca, and INFN, sezione di Milano–Bicocca,
Piazza della Scienza 3, I-20126 Milano, Italy}

\author{Harvey B.\ Meyer} 
\affiliation{Helmholtz-Institut Mainz, Johannes Gutenberg-Universit\"at Mainz,
D-55099 Mainz, Germany}
 \affiliation{PRISMA$^+$ Cluster of Excellence  \& Institut f\"ur Kernphysik,
Johannes Gutenberg-Universit\"at Mainz, D-55099 Mainz, Germany}

\author{Aman Steinberg}
\affiliation{PRISMA$^+$ Cluster of Excellence  \& Institut f\"ur Kernphysik,
Johannes Gutenberg-Universit\"at Mainz, D-55099 Mainz, Germany}
\affiliation{Fakult\"at für Physik, Universit\"at Bielefeld, D-33615 Bielefeld, Germany}

\author{Arianna Toniato}
\affiliation{PRISMA$^+$ Cluster of Excellence \& Institut f\"ur Kernphysik,
 Johannes Gutenberg-Universit\"at  Mainz,  D-55099 Mainz, Germany}

 \email{meyerh@uni-mainz.de}

\date{\today}

\begin{abstract}
We calculate the thermal rate of real-photon production in the quark-gluon plasma
at a temperature of $T=254{\rm\,MeV}$ using lattice QCD.
The calculation is based on the difference between the spatially transverse and longitudinal
parts of the polarization tensor, which has the advantage of falling off rapidly at large frequencies.
% and the associated spectral function obeys a superconvergent sum rule that we exploit in our analysis.
We obtain this linear combination in the time-momentum representation from lattice QCD with 
two flavors of quarks in the continuum limit with a precision of about two parts per mille.
Applying a theoretically motivated fit ansatz for the associated spectral function,
we obtain values for the photon rate that are in line with 
QCD weak-coupling calculations; for photon momenta $ 1.0\leq k[{\rm GeV}]\leq 1.4$, our non-perturbative results
constrain the rate to be no larger than twice the weak-coupling prediction.
We also provide a physics interpretation of the electromagnetic spectral functions valid for all frequencies
and momenta.
\end{abstract}

% \pacs{11.15.Ha, 11.25.Pm, 12.38.Aw} 
% { 12.38.Gc, 12.38.Mh, 25.75.-q}
\maketitle

\section{Introduction}

Strongly interacting matter undergoes a phase transition at a
temperature of about 150\,MeV~\cite{Borsanyi:2010bp,Bazavov:2011nk,Bhattacharya:2014ara}. Below the transition, the thermal
medium is characterized by hadrons (nucleons, pions, kaons, \dots) as
primary degrees of freedom, while well above the transition it is
characterized by quarks and gluons, the elementary degrees of freedom
of quantum chromodynamics (QCD).  The high-temperature phase, the
quark-gluon plasma (QGP), is probed experimentally in high-energy
heavy-ion collisions at $T\lesssim
500{\rm\,MeV}$~\cite{Braun-Munzinger:2015hba}.  One of the remarkable
properties of the medium is its ability to exhibit collective effects
in spite of the rapid expansion occurring in heavy-ion collisions. The
most prominent such effect is the large anisotropic flow observed in
heavy-ion collisions at RHIC and the LHC, pointing to a small shear
viscosity to entropy density ratio of the medium; see e.g.\ \cite{Shen:2015msa} and Refs.\ therein.  
In addition, probes of the medium that do not interact strongly are of great interest, since they
escape largely unscathed once produced.  In particular, the rate at
which photons are emitted by the QGP is a classic -- though challenging -- 
observable in heavy-ion experiments.
Direct photons with a transverse momentum below 2\,GeV
are found to admit an exponential spectrum, 
and models assuming the formation of the QGP are consistent with these measurements~\cite{Adare:2014fwh,Adam:2015lda}.
The production of weakly interacting particles by the QGP is
also an important issue in early-universe cosmology, for instance in
models which propose a keV-scale sterile neutrino as a
dark matter candidate~\cite{Asaka:2006rw,Asaka:2006nq}.

In this Letter we address the rate of photon emission from the QGP via
lattice QCD simulations. One motivation to perform the calculation is
that the rate vanishes in the limit of non-interacting quarks and
gluons; therefore it is a measure of the strength of their
interactions. Secondly, direct photons emitted in heavy-ion collisions
have been found to exhibit an unexpectedly large central value of
elliptic flow \cite{Adare:2015lcd,Acharya:2018bdy} -- albeit with
significant uncertainty, therefore addressing their thermal production
rate non-perturbatively can contribute to resolving the issue.
Thirdly, a controlled calculation of the photon rate paves the way for
calculating the production of other particles, such as lepton pairs --
relevant in heavy-ion phenomenology -- or sterile neutrinos --
relevant for validating or ruling out a dark matter candidate.

The main computational difficulty stems from the production of
weakly-interacting particles being a real-time process, which is
accessible from the Matsubara path integral formalism implemented in
lattice QCD only via an analytic continuation~\cite{Meyer:2011gj}.
Numerically, the latter amounts to a poorly conditioned inverse
problem discussed below.

\section{Theory background}

We consider the full set of spectral functions of the electromagnetic current\footnote{The Minkowski-space Dirac matrices satisfy $\{\gamma^\mu,\gamma^\nu\} = 2g^{\mu\nu}$ with $g^{\mu\nu}={\rm diag}(1,-1,-1,-1)$. Also, time-evolution in Eq.\ (\ref{eq:rhomunu}) is Minkowskian, 
$V^\mu(t,\vec x)\equiv e^{-i(\vec P\cdot \vec x -Ht)} V^\mu(0,\vec 0) e^{i(\vec P\cdot \vec x -Ht)}$.} 
$V^\mu = \sum_{f=u,d,s,\dots} Q_f \bar\psi_f \gamma^\mu \psi_f$,
\be\la{eq:rhomunu}
\rho^{\mu\nu}(\omega,\vec k) =  \!\!\int\! d^4x 
\, e^{i(\omega x^0 -\vec k\cdot \vec x)} \,
  {\rm Tr}\Big\{\frac{e^{-\beta H}}{Z(\beta)} \Big[V^\mu(x),V^\nu(0)^\dagger\Big] \Big\}.
\ee
For any four-vector $u^\mu$,
the form $u_\mu^\dagger\,\rho^{\mu\nu}(\omega,\vec k)u_\nu/\omega$ is real and non-negative;
for  $u^\mu$ real, it is also even in $\omega$.
Current conservation leads to
$\omega^2 \rho^{00}(\omega,\vec k) = k^i k^j \rho^{ij}(\omega,\vec k)$,
implying that\footnote{We use the notation $k \equiv |\vec k|$ and $\hat k^i = \frac{k^i}{k}$.} $(\hat k^i\hat k^j\rho^{ij} - \rho^{00})/\omega$ has the same sign as 
${\cal K}^2\equiv\omega^2-k^2$, and that it vanishes at lightlike kinematics, ${\cal K}^2=0$.
It will be useful to consider the linear combination
\be
\rho(\omega, k,\lambda) = (\delta^{ij}-\hat k^i \hat k^j)\rho^{ij} +\lambda\, (\hat k^i \hat k^j\rho^{ij}-\rho^{00}).
\ee
Defining the Euclidean correlator\footnote{The Euclidean current is defined by 
$V^{_{\rm E}}_\mu\equiv \sum_{f} Q_f\bar\psi_f\gamma_\mu^{_{\rm E}}\psi_f$, 
with $\{\gamma_\mu^{_{\rm E}},\gamma_\nu^{_{\rm E}}\}=2\delta_{\mu\nu}$. Also, time-evolution is Euclidean in Eq.\ (\ref{eq:GEmunu}),
$V^{_{\rm E}}_\mu(x) =  e^{x_0H-i\vec P\cdot \vec x} V^{_{\rm E}}_\mu(0) e^{-x_0H+i\vec P\cdot \vec x}$.}
\be\la{eq:GEmunu}
G^E_{\mu\nu}(x_0,\vec k) = \int d^3x\; e^{-i\vec k\cdot \vec x} \Big\< V^{_{\rm E}}_\mu(x)\, V^{_{\rm E}}_\nu(0)^\dagger\Big\>,
\ee
the corresponding linear combination
\be
G(x_0,k,\lambda) = (\delta^{ij}-\hat k^i \hat k^j)G^E_{ij} +\lambda\, (\hat k^i \hat k^jG^E_{ij}-G^E_{00})
\ee
admits the spectral representation
\be\la{eq:GmunuRho}
G(x_0,k,\lambda) = \int_0^\infty \frac{d\omega}{2\pi}\; \rho(\omega,k,\lambda)
\,\frac{\cosh[\omega(\beta/2-x_0)]}{\sinh(\beta\omega/2)}.
\ee

The production rate of dileptons with invariant mass-squared equal to ${\cal K}^2$, which occurs via a timelike photon,
is proportional to $\rho(\omega, k,1)$~\cite{McLerran:1984ay}.
To leading order in the fine-structure constant $\alpha = {e^2}/(4\pi)$,
the differential photon rate per unit volume of plasma can be written as 
\be\la{eq:photonrate}
{d\Gamma(k)} = e^2 \;\frac{d^3k}{(2\pi)^3\,2k}\; \frac{\rho(k,k,\lambda)}{e^{\beta k}-1}
\ee
and does not depend on $\lambda$. 
The Euclidean correlator $G(x_0,k,\lambda)$ probes the spectral function 
for all virtualities $ {\cal K}^2 \geq -k^2$. It is therefore desirable to have 
an interpretation of the spectral function for negative virtualities.
The cross-section per unit volume for an electron scattering on the medium through the exchange of a spacelike photon
is given by
\ba\la{eq:Xsec}
\frac{d^2\sigma}{L^3dp^0{}'d\Omega} &=& \frac{e^4(p^0{}'/p^0)}{32\pi^3{\cal K}^4}  \ell_{\mu\nu} 
\frac{\rho^{\mu\nu}(k^0,\vec k)}{1-e^{-\beta k^0}},
\\
\ell^{\mu\nu} &\equiv & 2(p^\mu p'{}^\nu + p^\nu p'{}^\mu - g^{\mu\nu} (p\cdot p')),
\nonumber
\ea
with $p$ and $p'$ respectively the initial and final electron momenta and $k=p-p'$.
Eq.\ (\ref{eq:Xsec}) refers to the rest-frame of the thermal medium.
More generally, the vector spectral functions can be interpreted as the ability 
of the medium to dissipate the energy stored in electromagnetic fields:
consider coupling the plasma to a harmonic external vector potential 
$\vec A(t,\vec x) = {\rm Re}(\vec A_{\vec k} e^{i(\vec k\cdot \vec x-\omega t)})$,
by adding the term $\Delta H = -e\int d^3x \; \vec j\cdot \vec A$ to the Hamiltonian.
The energy of the external electromagnetic fields is given by  $E_{\rm e.m.} = \frac{1}{2}\int d^3x\,(\vec E^2+\vec B^2)$.
A fraction of this energy gets transferred to the medium per unit time and turned into heat.
We find, for the transverse and longitudinal cases, the following rates of energy transfer,
\ba
\vec A_{\vec k}\perp \vec k\;:\;&& 
\frac{-1}{E_{\rm e.m.}}\frac{dE_{\rm e.m.}}{dt} =  e^2 \frac{\omega\, (\delta^{ij}-\hat k^i \hat k^j)\rho^{ij}(\omega,\vec k)}{2(\omega^2+k^2)} ,
\\
\vec A_{\vec k}\,\|\,\vec k\;:\;&& \frac{-1}{E_{\rm e.m.}}\frac{dE_{\rm e.m.}}{dt} = e^2\frac{\hat k^i \hat k^j\rho^{ij}(\omega,\vec k)}{\omega}
  = e^2\frac{\omega}{ k^2} \rho^{00}(\omega,\vec k).  
\nonumber
\ea
These equations provide an interpretation of the spectral functions for all virtualities.
The positivity of the spectral functions on the right-hand side guarantees that the medium obeys the second law of thermodynamics.

Given the goal of computing the photon rate, computationally
its non-dependence on the value of the parameter $\lambda$ can be exploited to one's advantage.
We choose $\lambda=-2$, because as a combined consequence of current conservation and Lorentz invariance,
$\rho(\omega, k,-2)$ vanishes identically in the vacuum (at zero temperature).
Due to the latter property and because $\rho(\omega,k=0,-2)$ vanishes exactly for $\omega\neq 0$ due to 
charge conservation, we expect from the operator-product expansion
\be\la{eq:asymp}
 \rho(\omega,k,-2) \propto k^2/\omega^4, \qquad \omega\gg \pi T,k.
\ee 
This strong suppression in the ultraviolet implies a superconvergent sum rule for $\rho(\omega,k,-2)$,
\be\la{eq:SR}
\int_0^\infty d\omega\; \omega\;\rho(\omega,k,-2) = 0.
\ee
Spectral positivity implies that $\rho(\omega,k,-2)/\omega$ is non-negative for ${\cal K}^2<0$, and it must become negative 
for  ${\cal K}^2>0$ in order to satisfy the sum rule (\ref{eq:SR}).

There are two regimes in which the functional form of the spectral function is known.
In the infrared limit, the $\rho^{00}$ contribution 
parametrically dominates $\rho(\omega,k,-2)$ and the hydrodynamic prediction is 
\be\la{eq:rhom2hydro}
\rho(\omega,k,-2)/\omega \approx \frac{4\chi_s\, Dk^2}{\omega^2+(Dk^2)^2}
\qquad \omega,k\ll D^{-1},
\ee
where $D$ is the diffusion coefficient and $\chi_s \equiv \beta\,G^{00}(x_0,\vec 0)$
the static susceptibility.
Therefore, following~\cite{Ghiglieri:2016tvj}, we define the effective diffusion coefficient 
\be\la{eq:Deff}
D_{\rm eff}(k) \equiv \frac{\rho(\omega=k,k,\lambda)}{4\chi_s k},
\ee
which is proportional to the photon rate and 
tends to $D$  in the limit $k\to0$.
In the weak-coupling regime, results at order $g^2$ have recently become available 
for general $(\omega,k)$~\cite{Laine:2013vma,Jackson:2019yao}.
The photon rate itself has been obtained at order $g^3$ in~\cite{Ghiglieri:2013gia}.

From here on we set $\lambda=-2$ and omit the last argument of $\rho(\omega,k,\lambda)$ and $G(\omega,k,\lambda)$.  

\section{The lattice calculation}

\begin{table}
\begin{tabular}{|c|c|c|c|c|c|}
\hline
label & $(6/g_0^2,\kappa)$ & $1/(aT)$ & $N_{\rm conf}$ & $\frac{\rm MDUs}{\rm conf\phantom{_A}}$  & $\tau_{\rm int}[Q^2(\bar t)]$ \\
\hline
F7&$(5.3,0.13638)$          & 12          & 482  & 20 & 11.3(15) \\  
O7&$(5.5,0.13671)$          & 16          & 305  & 20 & 19(5) \\ 
W7 & $(5.685727,0.136684)$  & 20          & 1566 &  8 & 81(23) \\ 
X7&$(5.827160,0.136544)$    & 24          & 511  & 10 & 490(230) \\ 
\hline
\end{tabular}
\caption{\la{tab:sim} Simulations at a fixed temperature of $T=(254\pm5)\,$MeV and fixed aspect ratio $TL=4$.
For orientation, the transition temperature is about 211\,MeV~\cite{Brandt:2016daq}.
The number of point sources per configuration is 16 in all cases. The autocorrelation time of the squared topological charge defined 
at gradient-flow time~\cite{Luscher:2010iy}
$\bar t=\beta^2/80$ is given in molecular-dynamics units (MDUs).}
\end{table}

We use lattice QCD with an isospin doublet of O($a$)
improved Wilson fermions at a temperature of $T=254{\rm\,MeV}$; the
details of the lattice action can be found in~\cite{Fritzsch:2012wq}
and references therein.  Table \ref{tab:sim} lists our ensembles,
which allow us to take the continuum limit at a fixed temperature.
All but the finest ensemble have a renormalized quark mass of $m^{\rm
  \overline{MS}}\simeq 13\,{\rm MeV}$ in the ${\rm \overline{MS}}$
scheme at a renormalization scale of $\mu=2{\rm\,GeV}$; on the finest
ensemble, we have $m^{\rm \overline{MS}}\simeq 16\,{\rm MeV}$.
Quark-mass effects, which are suppressed by $(m/T)^2$ in the chirally
symmetric phase, are therefore expected to be negligible. 
The ensembles F7, O7 and X7 were generated using the MP-HMC
algorithm~\cite{Hasenbusch:2001ne} in the implementation described in
Ref.~\cite{Marinkovic:2010eg} based on the DD-HMC
package~\cite{CLScode}, while 
ensemble W7 was generated using twisted-mass Hasenbusch frequency splitting 
in the version 1.6 of openQCD~\cite{Luscher:2012av,CLScode21}.
The ensembles labelled F7 and O7 have bare
parameters identical to the zero-temperature F7 and O7 ensembles
described in~\cite{Fritzsch:2012wq}, for which the pion mass is
269\,MeV.

\begin{figure}
   \centering
\includegraphics[width=1.04\columnwidth]{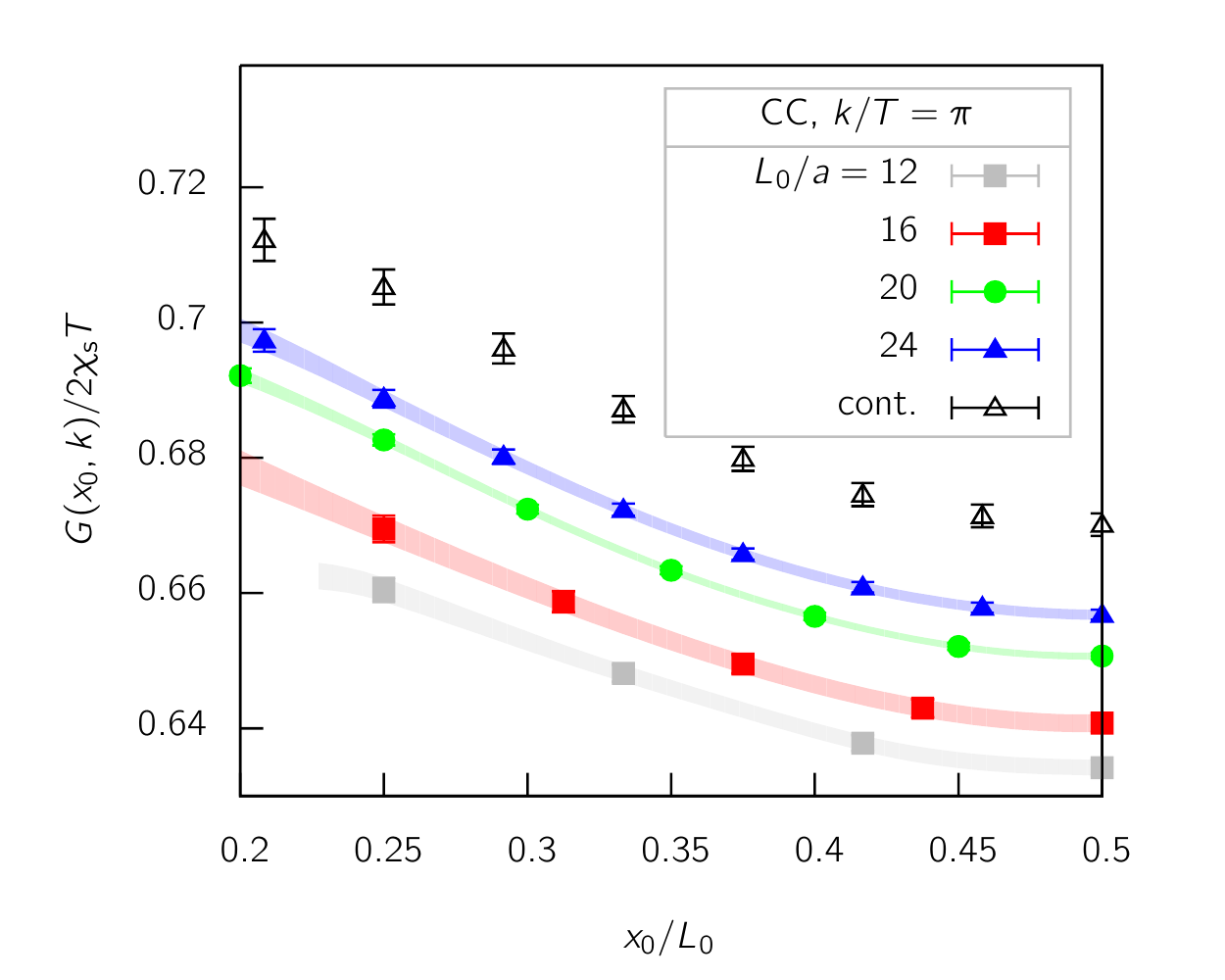}
\caption{The treelevel-improved correlator $G(x_0,k=\pi T)$ obtained at
different lattice spacings and its continuum limit. The latter is obtained by 
jointly extrapolating four discretizations of $G(x_0,k)$ from $1/(aT)=16$, 20 and 24 to the continuum.}
   \label{fig:contlim}
\end{figure}

We compute the correlator $G(x_0,k)$ of the isovector current
$\frac{1}{\sqrt{2}}\bar\psi \gamma_\mu \tau^3\psi$, which consists of a
single connected Wick contraction\footnote{
In order to keep the notation concise, we do not explicitly distinguish between the quantities derived from the isovector and from the electromagnetic current.
To obtain the photon rate from our results for $D_{\rm eff}(k)$, we recommend
using Eq.\ (\ref{eq:photonrate}) with $\rho(k,k)= 4k D_{\rm eff}(k)\cdot \chi_s[Q_f]$ in the approximation
$\chi_s[Q_f] \simeq  C_{\rm em}\cdot \chi_s[{\rm isovector}]$,   with $C_{\rm em} = \sum_{f=u,d,s} Q_f^2=2/3$.}.  
The corresponding static susceptibility amounts
to $G(x_0,0)/(2T^3) = \chi_s/T^2 = 0.880 (9)_{\rm stat} (8)_{\rm syst}$ in the continuum
limit, where the systematic error reflects the dependence on using
different prescriptions for the renormalisation of the local vector
current.
We employ the local and the conserved vector currents, resulting in four discretizations of
$G(x_0,k)$, and perform a constrained simultaneous continuum
extrapolation.  We have computed the leading-order perturbative
lattice predictions, so that we are able to correct for the
corresponding cutoff effects affecting our Monte-Carlo data. To avoid
incurring large cutoff effects at short distances, we omit data points
for $x_0<x_0^{\rm min}$, where $x_0^{\rm min}=\beta/4$ is our default
value. We thus have data points for $G(x_0,k)$ at $x_0^{(i)}
=\frac{i}{24}\cdot\beta$, $i\in\{6,7,8,9,10,11,12\}$.  Given the high
accuracy of the data, we are led to leave out the ensemble with the
coarsest lattice spacing from the continuum limit. Figure
\ref{fig:contlim} illustrates the correlator obtained at different
lattice spacings and its continuum limit.  The relative statistical
precision of the continuum correlator is one to two permille.  It is
well-known that the topological charge $Q$ acquires a long
autocorrelation time at small lattice spacings, and our simulations
confirm this effect. However, we have found the dependence of the
vector correlator of interest on $|Q|$ to be at most at the 3\% level.
Therefore the vector correlator only suffers a modest increase in uncertainty
from this algorithmic difficulty.

\begin{figure}[t]
   \centering
\includegraphics[width=1.04\columnwidth]{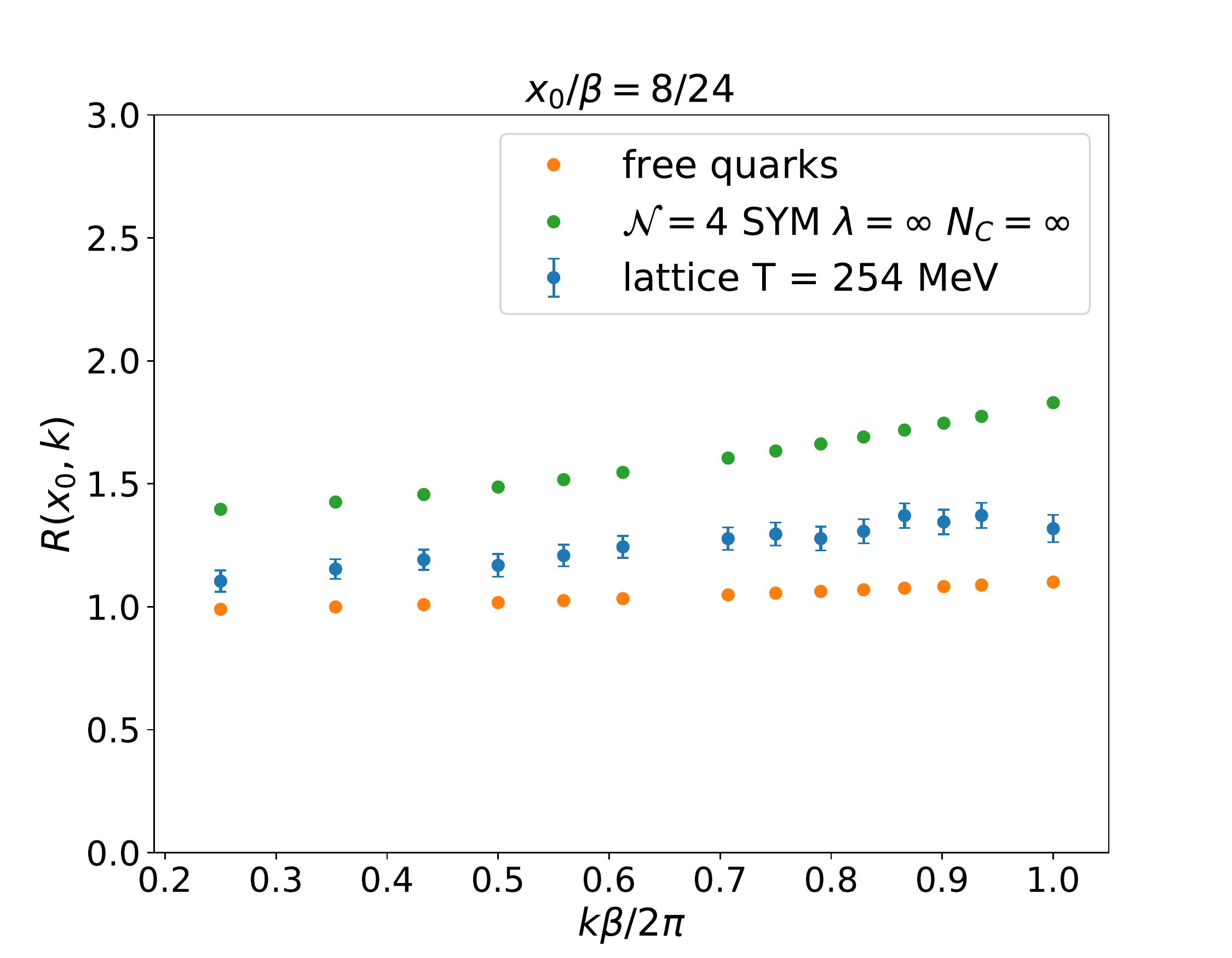}
\caption{The observable $R(x_0,k)$ (see Eq.\ (\ref{eq:R0k})) for $x_0=\beta/3$ in $N_{\rm f}=2$ QCD at $T=254\,$MeV,
compared to its prediction for non-interacting fermions and for the strongly coupled SYM theory.}
   \label{fig:newobs}
\end{figure}

We define the observable
\be\la{eq:R0k}
R(x_0,k) \equiv \frac{16\pi}{(\beta - 2x_0)^2k^2}\; \Big[\frac{G(x_0,k)}{G(\beta/2,k)}-1\Big].
\ee
Expressed in terms of the spectral function, in the limit $x_0\to
\beta/2$ it describes the ratio of the $\omega^2$ moment to the
$\omega^0$ moment of $\rho(\omega,k)/\sinh(\omega\beta/2)$.  The
$1/k^2$ factor allows for a finite $k\to 0$ limit. It is instructive to
compare the results from lattice QCD with the theory of non-interacting quarks
as well as with an extreme opposite, namely the ${\cal N}=4$ super-Yang-Mills (SYM) theory in the limit of 
infinite 't Hooft coupling and infinite number of colors; 
the spectral functions of the latter are obtained via the AdS/CFT correspondence~\cite{CaronHuot:2006te}.
One qualitative difference between the spectral functions of the strongly coupled SYM
theory and of free quarks is that in the former case the positive
spectral weight of the spacelike region $\omega^2<k^2$ `leaks' into
the timelike region; see especially the second panel of Fig.\ \ref{fig:SF}. For
$k\approx \pi T$, this feature results in the observable $R(x_0,k)$
being about 1.5 times larger in the former theory. It is thus
interesting to ask how $R(x_0,k)$ behaves in QCD at the temperature of
254\,MeV.  The observable is displayed in Fig.\ \ref{fig:newobs}. The
QCD values lie less than 20\% above the non-interacting values.

\section{Analysis of the spectral function}

To obtain a global picture of the spectral function without committing
to any specific functional form, in~\cite{Brandt:2017vgl} we applied
the Backus-Gilbert method to our data. The results confirm the
theoretical expectation that most of the spectral weight is contained in the
spacelike region $\omega^2<k^2$.

\begin{figure}[t!]
\centering
\!\!\!\!\!\!\includegraphics[width=1.04\columnwidth]{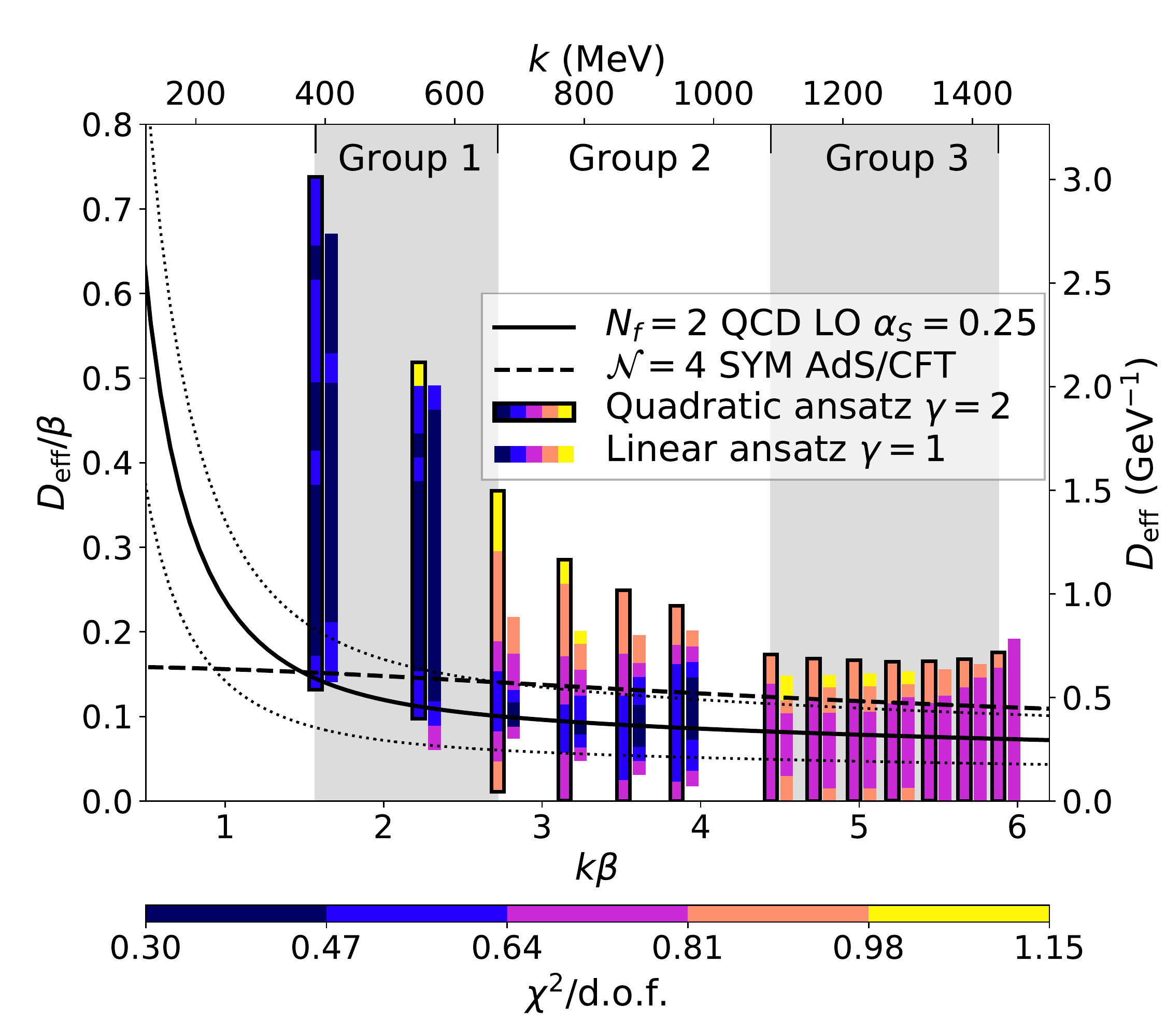}
\caption{Lattice results for the effective diffusion coefficient $\Deff(k)$, defined by Eqs.\ (\ref{eq:photonrate}) and (\ref{eq:Deff}).
The color-coded vertical bars represent those values of $\Deff$ for which a spectral function of the form (\ref{eq:ansatz}) exists 
that has a $p$-value above 0.32. The colors indicate the smallest $\chi^2/{\rm d.o.f.}$ found for a given value of $\Deff$.
Shaded areas identify the momentum groups that are fitted simultaneously; for each momentum, results are shown both for the $\gamma=1$ and $\gamma=2$
parametrizations of the $k$-dependence of the nonlinear parameters.
Analytical results from perturbative QCD~\cite{Arnold:2001ms} and from the strong-coupling limit of ${\cal N} = 4$ 
super-Yang-Mills theory~\cite{CaronHuot:2006te} are shown for comparison.}
   \label{fig:Deff}
 \end{figure}

A second method~\cite{Brandt:2017vgl}, which we now pursue further,
consists in applying an explicit fit ansatz for the spectral function,
\be\la{eq:ansatz}
\rho(\omega,k) = 
\frac{A(1+ B\omega^2)\; \tanh(\omega\beta/2)}{[(\omega-\omega_0)^2+b^2][(\omega+\omega_0)^2+b^2][\omega^2+a^2]}.\quad 
\ee
The ansatz satisfies the expected large-$\omega$ behavior (\ref{eq:asymp}).
We always determine the parameter $ B$ in terms of $(\omega_0,a,b)$ by imposing the sum rule (\ref{eq:SR})
and require $B\geq -1/k^2$ to satisfy the spectral positivity condition for $\omega^2<k^2$.
Thus, for a single momentum $k$, Eq.\ (\ref{eq:ansatz}) amounts to a four-parameter fit.
The Euclidean correlator resulting from the spectral function (\ref{eq:ansatz}) can be expressed
as a linear combination of Lerch transcendents $\Phi(e^{\pm2\pi i x_0},1,\frac{1}{2}+i\frac{\omega_p}{2\pi})$, 
where $\omega_{\rm p}$ are the frequency poles of $\rho(\omega,k)/\tanh(\omega\beta/2)$.

\begin{figure}[t!]
   \centering
\includegraphics[width = 1.0\columnwidth]{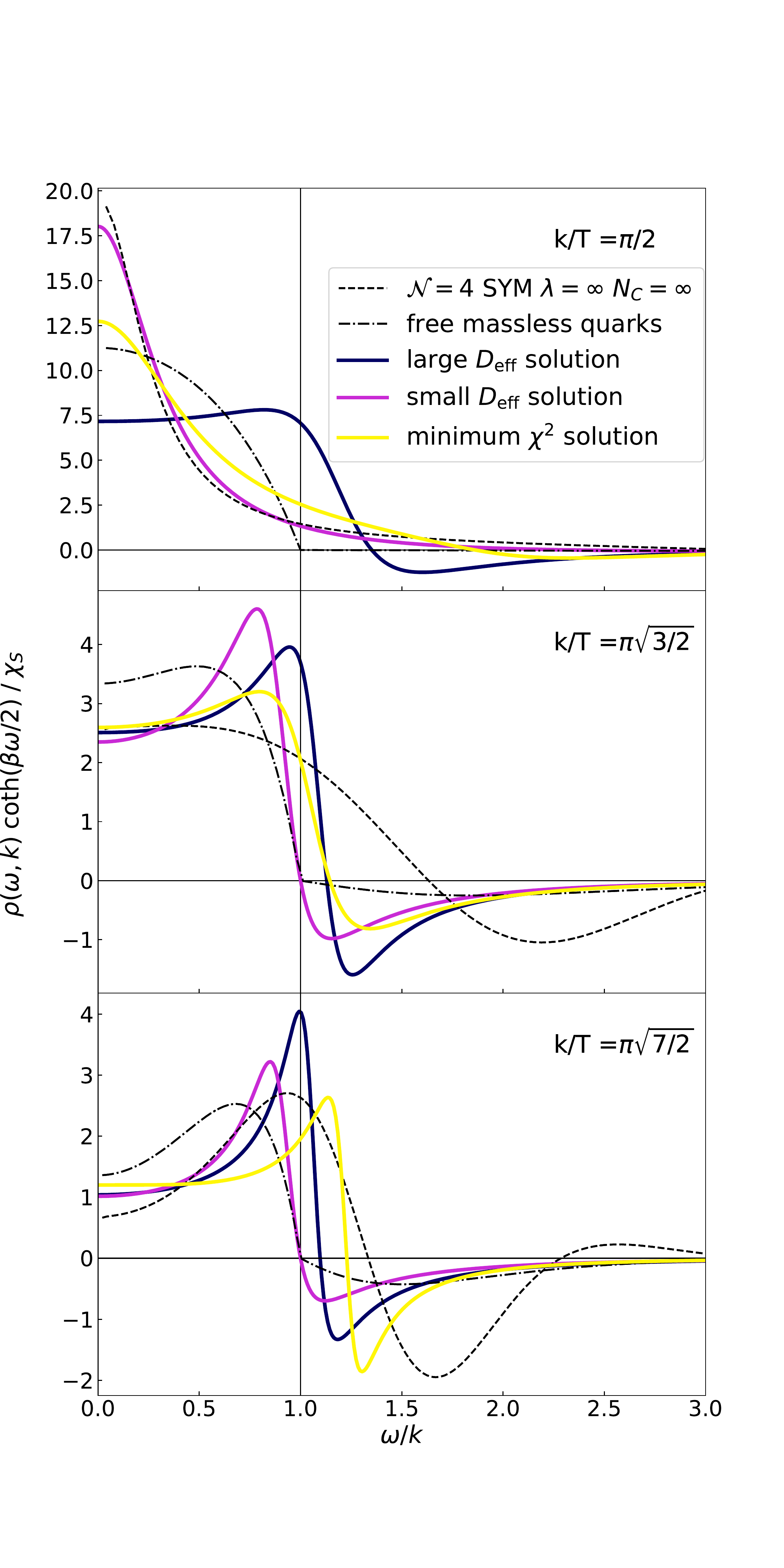}
   \caption{Representative spectral functions obtained from lattice QCD data for three different spatial momenta.
They are compared to the spectral functions of non-interacting quarks and of the strongly coupled SYM theory.}
\label{fig:SF}
\end{figure}

We impose the following physically motivated constraints on the parameters.
Spectral positivity implies $A\geq0$ and  $B\geq -1/k^{2}$.
Furthermore, since there cannot be arbitrarily long relaxation times in the system,
we impose the condition
\be
{\rm Im}(\omega_{\rm p}) >{\rm min}(D_{\rm strong}k^2, D_{\rm weak}^{-1})
\ee
on the poles, where $D_{\rm strong} =
\frac{1}{2\pi T}$ is the diffusion coefficient of the strongly coupled SYM theory 
and  $D_{\rm weak}^{-1}$  the inverse QCD
diffusion coefficient at leading-order in the perturbative expansion, which
we set to $0.46T$ based on the results of~\cite{Arnold:2003zc}.
This condition reflects the fact that $Dk^2$ is the rate of dissipation 
of a perturbation in the charge density, while $D^{-1}$ provides an estimate of 
the relaxation rate of a homogeneous current.

In order to increase the discriminative power of our fits, we 
simultaneously fit data at different momenta. 
The correlators have been computed for all spatial momenta $\vec k =
\frac{\pi\,T}{2} \, \vec \nu$ for $\vec \nu\in\mathbb{Z}^3$ and $n\equiv |\vec \nu|^2
\leq 16$.  We found it convenient to split the set of available momenta
into three groups, $1\leq n \leq 3$, $3\leq n \leq 8$ and
$8\leq n \leq 14$, which contain respectively $N_k=3$, 5 and 7
momentum values. 
The number of data points entering a fit is thus
given by $N_k N_t$, the number $N_t$ of 
Euclidean times being seven in our data set.
Within a momentum group, we parameterize  the 
momentum dependence of ansatz (\ref{eq:ansatz}) by expressing the
nonlinear parameters $a$, $b$ and $\omega_0$ as functions of the
momentum. We consider two polynomial forms in our analysis,
\be
\label{Q}
\left(\begin{array}{c}  a(k) \\ b(k) \\ \omega_0(k) \end{array}\right)
= \left(\begin{array}{c}  a_0 \\ b_0 \\ W_0 \end{array}\right)
+ (k^\gamma-k^\gamma_{\rm min})\left(\begin{array}{c}  a_2 \\ b_2 \\ W_2 \end{array}\right),
\ee
with $\gamma=1$ or 2 and where $k_{\rm min}$ is the smallest momentum in the group.

Since the covariance matrix $C$ of the data points is sizeable, 
we have used the regularized matrix $\tilde C$, constructed according to\footnote{No summation convention is applied in Eq.\ (\ref{eq:Creg})} 
\ba\la{eq:Creg}
\tilde C^{(nn')}_{x_0x_0'} &=& (1-y) \delta^{nn'} \hat C^{(nn)}_{x_0x_0'} + y\, \hat C^{(nn')}_{x_0x_0'},
\\
\hat C^{(nn')}_{x_0x_0'} &=& (1-x) \delta_{x_0x_0'} C_{x_0x_0}^{(nn')} + x\, C_{x_0x_0'}^{(nn')} .
\nonumber
\ea
We have studied the stability of our results with respect to the
regularization parameters $x,y$ and found little dependence on them
around the values we chose~\cite{Brandt:2019shg}.  For instance, both
$x$ and $y$ were set to 0.95 for the second momentum group.  The
importance of preserving correlations among the input data points when
addressing the inverse problem has been emphasized
previously~\cite{Ding:2016hua}.

For each momentum group, we performed a scan in the six-dimensional
space of non-linear parameters ($a_0$,$a_2$,$b_0$,$b_2$,$W_0$,$W_2$),
while, at each momentum, the parameter $B$ is determined by imposing
the sum rule (\ref{eq:SR}) and the linear parameter $A$ by minimizing
the $\chi^2$.  The number of fit parameters is thus given by $6 +
N_k$, and the number of degrees of freedom for each of the three
momentum groups is 12, 24 and 36 respectively.  We calculate the
$p$-value of each set of parameter values and consider that it
provides a satisfactory description of the correlator whenever
$p>0.32$. If the condition is satisfied, the corresponding
$D_{\rm eff}(k)$ are marked as being compatible with the lattice data,
and the associated $p$-value is recorded.

Before describing our results for $D_{\rm eff}(k)$,
we briefly present the outcome of our procedure when applied
to mock Euclidean data generated from known spectral functions.  For
these tests, we have used the spectral functions of non-interacting
quarks as well as those of the strongly coupled SYM theory. In order to be realistic,
we re-use the covariance matrix of our lattice QCD data, rescaled
so as to achieve the same relative error on the correlator.
In both cases, we find that the correct value of $D_{\rm eff}(k)$
is one of those having a $p$-value above 0.32. The output spectral functions
yielding the highest $p$-value tend to have a somewhat larger value of $D_{\rm eff}(k)$.

Our final results for the $D_{\rm eff}(k)$ values yielding a $p$-value
above 0.32 for the QCD correlator at $T=254\,$MeV are displayed in
Fig.\ \ref{fig:Deff}. We show results for both the linear and the
quadratic dependence on $k$, $\gamma=1$ and 2. We observe that for the third momentum
group, containing momenta above 1.0\,GeV, the values of $D_{\rm
  eff}(k)\cdot {\rm GeV}$ cover the interval $[0,\,0.7]$ and are thus
compatible both with the leading-order weak-coupling prediction~\cite{Arnold:2001ms} and
the strongly-coupled SYM prediction~\cite{CaronHuot:2006te}, which lie between 0.3 and 0.5.
Moreover, the weak-coupling prediction is among those values with the
highest $p$-value.  In the second momentum group, the range of
acceptable $D_{\rm eff}(k)$ values covers a range up to about twice
the strongly-coupled SYM value (for the ansatz quadratic in $k$),
while again the weak-coupling prediction has one of the highest
$p$-values. In the group of smallest momenta, the lattice data loses
sensitivity to the photon rate.  Particularly, the data does not
exclude large values of $D_{\rm eff}(k)$.  Finally, we remark that our
fits yield a strong correlation between the values of $D_{\rm eff}(k)$
at successive $k$~\cite{Brandt:2019shg}.

It is instructive to look at the full frequency dependence of the
spectral functions which describe the QCD correlators. In
Fig.\ \ref{fig:SF}, as representative examples for the three spatial
momenta $k=(0.40,0.98,1.49)\,{\rm GeV}$, we show spectral functions
that correspond to the upper and lower end of the $D_{\rm eff}(k)$
ranges shown in Fig.\;\ref{fig:Deff}. We also display the spectral
function leading to the smallest $\chi^2$, and for comparison, the
spectral functions of non-interacting quarks as well as those of the
strongly coupled SYM theory. For the second and third momenta, we
observe that all three spectral functions describing the QCD
correlators exhibit a smooth behaviour for $\omega^2<k^2$ and admit a
maximum near the point $\omega=k$, its precise location being
tightly linked to the value of $D_{\rm eff}(k)$ and hence to the photon
emission rate.

\section{Conclusion}

Using lattice simulations in the quark-gluon plasma phase of
QCD with two dynamical quark flavors at a temperature of 254\,MeV, we have computed one
particularly ultraviolet-soft component of the polarization tensor in
the continuum limit. This component determines the photon emission
rate from the medium via analytic continuation, in practice however
one is faced with the inverse problem (Eq.\ (\ref{eq:GmunuRho})) for the spectral
function.  We explored exhaustively the parameter space of the
Pad\'e-form spectral functions in Eq.\ (\ref{eq:ansatz}).  The photon rate is given,
up to kinematical factors, by the spectral function at photon
kinematics, $\omega=k$, and normalizing this quantity by the
well-determined static charge susceptibility, one obtains the
effective (momentum-dependent) diffusion coefficient.  Within the
explored family of spectral functions, we determined which values of
this coefficient are compatible with the Euclidean data; our result is
displayed in Fig.\ \ref{fig:Deff}.  We have validated our handling of
the inverse problem by applying the same procedure to two field
theories that represent extreme opposite cari\-ca\-tures of the
quark-gluon plasma.  Our results imply non-perturbative constraints on
the possible rate of photon emission from the QGP at a temperature
typical for the strongly interacting system created in heavy-ion
collision experiments.  We largely confirm the weak-coupling
predictions, in spite of the relatively low temperature of 254\,MeV.
Our results are also in good agreement with those of a previous lattice calculation
performed in the quenched approximation~\cite{Ghiglieri:2016tvj}.

As a study based on correlators in the theory of non-interacting
quarks shows, adding data points at shorter Euclidean time
significantly enhances the ability of the data to exclude large values
of $D_{\rm eff}$, particularly at low photon momenta.  This calls for
even finer lattices to be used, which represents a challenge in view
of the large lattice sizes required and the long associated
autocorrelation times.

As described in~\cite{Brandt:2017vgl}, an analogous combination of
correlators can be applied to energy-momentum tensor correlators to
extract an effective shear viscosity $\eta_{\rm eff}(k)$.  We finally
remark that a different strategy has also recently been proposed to
compute the photon rate using a dispersion relation at fixed,
vanishing virtuality~\cite{Meyer:2018xpt}. The ultraviolet-soft
channel employed here also plays an important role in the
implementation of this alternative method.

\acknowledgments{We thank B.B.\ Brandt and A.~Francis, who were involved in the early stages 
of this project~\cite{Brandt:2017vgl}, as well as M.\ Laine and G.D.\ Moore for discussions and encouragement. 
This work was supported in part by DFG Grant ME 3622/2-2 and by the
European Research Council (ERC) under the European Union’s Horizon
2020 research and innovation program through Grant Agreement
No.\ 771971-SIMDAMA.  
The work of M.C.\ is supported by the European Union’s Horizon 2020 research and innovation programme 
under the Marie Sklodowska-Curie grant agreement No.\ 843134-multiQCD.
A.S.\ was supported in part by DFG - project number 315477589 - TRR 211.
The generation of gauge configurations as well as
the computation of correlators was performed on the Clover and Himster2 platforms
at Helmholtz-Institut Mainz and on Mogon II at Johannes Gutenberg University Mainz.
We have also benefitted from computing resources at Forschungszentrum J\"ulich allocated under NIC project HMZ21.}

%%%%%%%%%%%%%%%%%%%%%%%%%%%%%%%%%%%%%%%%%%%%%%%%%%%%%%%%%%%%%%%%
\bibliography{/Users/harvey/BIBLIO/viscobib}
%%%%%%%%%%%%%%%%%%%%%%%%%%%%%%%%%%%%%%%%%%%%%%%%%%%%%%%%%%%%%%%%

\end{document}